\documentclass{PoS}

\title{Charmless $B_s^0$ decays at Belle}

\ShortTitle{Charmless $B_s$ decays}

\author{\speaker{Bilas Pal}\thanks{On behalf of the Belle Collaboration}\\
        University of Cincinnati\\
        E-mail: \email{palbs@ucmail.uc.edu}}

\abstract{We report on recent results on charmless hadronic and radiative rare $B_s^0$ decays, based on the data collected by the  Belle experiment at the KEKB $e^+e^-$ collider. These include the studies of 
$B_s^0\rightarrow K^0\bar{K^0}$, $B_s^0\rightarrow\phi\gamma$ and $B_s^0\rightarrow\gamma\gamma$.
  }

\FullConference{XIII International Conference on Heavy Quarks and Leptons\\
		22-27 May, 2016\\
		Blacksburg, Virginia, USA}

\begin{document}

\section{Introduction}
Charmless hadronic $B$ decays are suppressed compared to other hadronic $B$ decays and hence can be excellent probes for new physics beyond the Standard Model (SM). In this paper, we present recent results from the Belle experiment on the charmless hadronic and radiative $B_s^0$ decays $B_s^0\rightarrow K^0\bar{K^0}$, $B_s^0\rightarrow\phi\gamma$ and $B_s^0\rightarrow\gamma\gamma$.

The main challenge of studying  the charmless $B_s^0$ decays is the suppression of overwhelmingly large background arising from continuum $e^+e^-\to q\bar{q}~(q=u,~d,~c,~s$) production.  To suppress this background, we use a multivariate analyzer
based on a neural network. The neural network uses the so-called event shape variables  to discriminate continuum events, which tend to be jetlike, from spherical $B\bar{B}$ events. Signal decays are identified by  two kinematical variables: the beam-energy-constrained mass $M_{\rm bc}= \sqrt{E^2_{\rm beam}-|\vec{p}^{}_{B}|^2c^2}/c^2$ and  the energy difference 
$\Delta E=E_{B}-E_{\rm beam}$.  To determine the signal yield, normally an unbinned extended maximum likelihood fit is applied to all candidate event using the above two  kinematical variables and other useful information. The signal probability density functions (PDF) of these two variables are typically studied from Monte Carlo (MC) simulation and the background PDF can be obtained  either from  MC simulation or sideband data. A high statistics control sample of similar topology is used  to understand potential data/MC differences.

\section{Observation of the decay $B_s^0\rightarrow K^0\bar{K^0}$}
\begin{figure}[htb]
\centering
\includegraphics[width=0.45\textwidth]{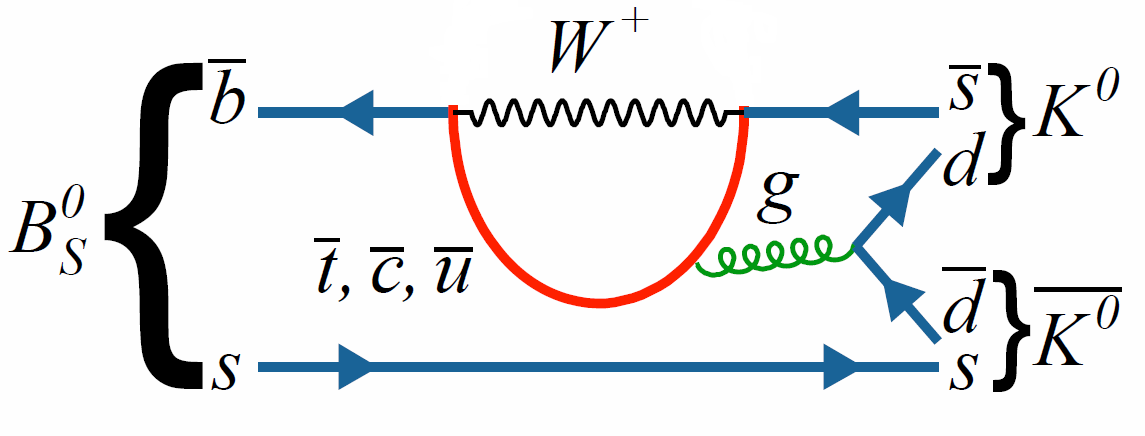}
\caption{\small Loop diagram for $B_s^0\rightarrow K^0\bar{K^0}$ decays.  }
\label{fig:feynman}
\end{figure}

The two-body decays $B_s^0\rightarrow h^+h'^-$, where $h^{\scriptscriptstyle(}\kern-1pt{}'\kern-1pt{}^{\scriptscriptstyle)}$ is
either a pion or kaon, have now all been observed~\cite{PDG}.
In contrast, the neutral-daughter decays $B_s^0\rightarrow h^0h'^0$ have
yet to be observed. The decay $B_s^0\rightarrow K^0\bar{K^0}$~\cite{charge-conjugate}
is of particular interest because the branching fraction is predicted
to be relatively large. In the SM, the decay
proceeds mainly via a $b\rightarrow s$ loop (or ``penguin") transition as shown
in Fig.~\ref{fig:feynman}, and the branching fraction is predicted
to be in the range $(16-27)\times10^{-6}$~\cite{SM-branching}.
The presence of non-SM particles or couplings could enhance 
this value~\cite{Chang:2013hba}. It has been pointed out
that $CP$ asymmetries in $B_s^0\rightarrow K^0\bar{K^0}$ decays are
promising observables in which to search for new
physics~\cite{susy}. 

The current upper limit on the branching fraction,
$\mathcal{B}(B_s^0\rightarrow K^0\bar{K^0})<6.6\times 10^{-5}$ at 90\% 
confidence level (C.L.), was set by the Belle Collaboration using 
$23.6~{\rm fb^{-1}}$ of data recorded at the
$\Upsilon(5S)$ resonance~\cite{Peng:2010ze}.
The analysis presented here uses the full data set of
$121.4~{\rm fb^{-1}}$ recorded at the~$\Upsilon(5S)$.
Improved tracking, $K^0$ reconstruction, and continuum suppression algorithms are also used in this analysis. 
The data set corresponds to $(6.53\pm 0.66)\times10^6$  $B_s^0\bar{B_s^0}$
pairs~\cite{Oswald:2015dma} produced in three $\Upsilon(5S)$ decay
channels: $B_s^0\bar{B_s^0}$, $B_s^{*0}\bar{B_s^0}$ or $B_s^0\bar{B}_s^{*0}$, and $B_s^{*0}\bar{B}_s^{*0}$.
The latter two channels dominate, with production fractions
of $f_{B_s^{*0}\bar{B_s^0}}=(7.3\pm1.4)\%$ and $f_{B_s^{*0}\bar{B}_s^{*0}}=(87.0\pm1.7)$\%~\cite{Esen:2012yz}.
The $B_s^{*0}$ decays via $B_s^{*0}\rightarrow B_s^0\gamma$, and the $\gamma$ is not reconstructed.

\begin{figure*}[t]
  \includegraphics[width=0.32\textwidth]{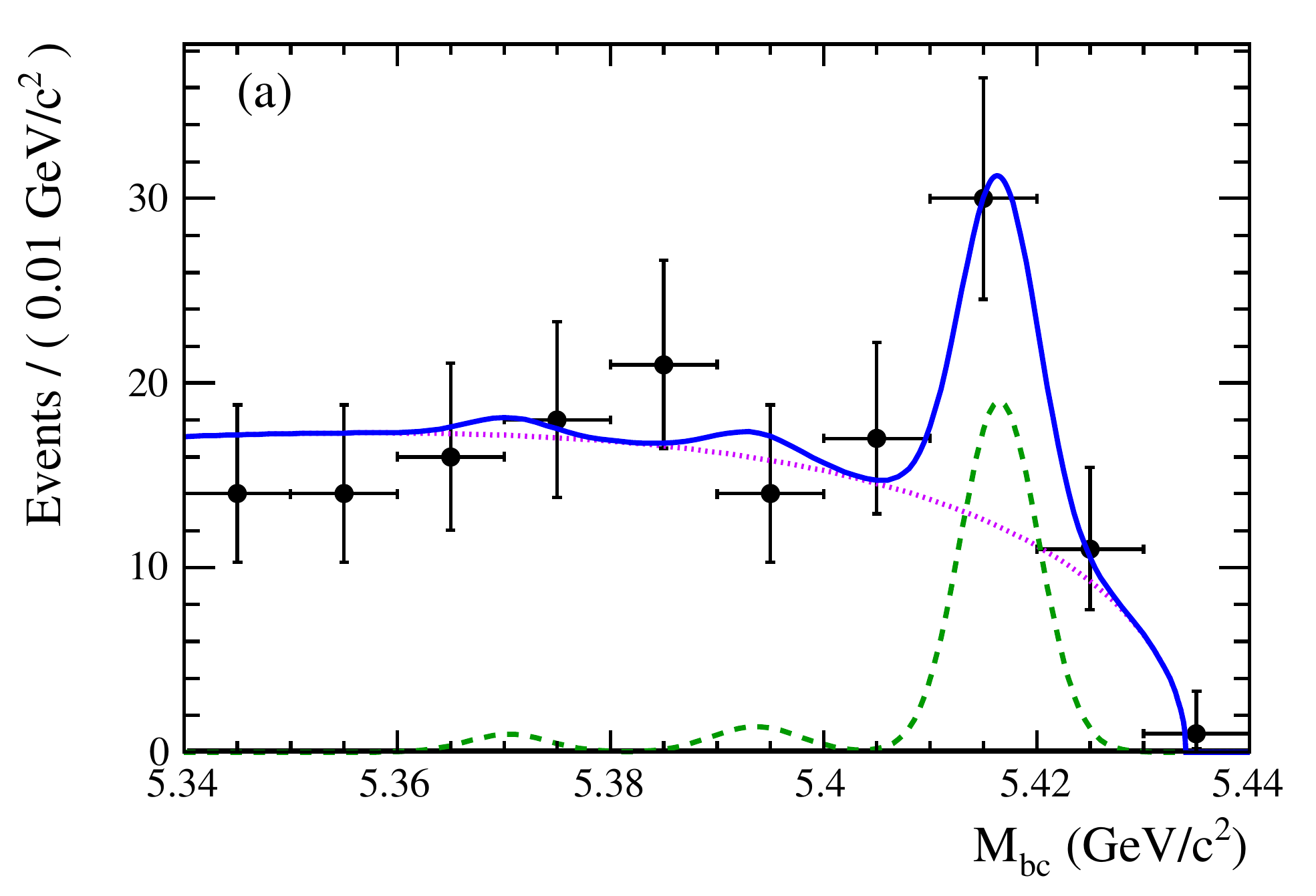}
  \includegraphics[width=0.32\textwidth]{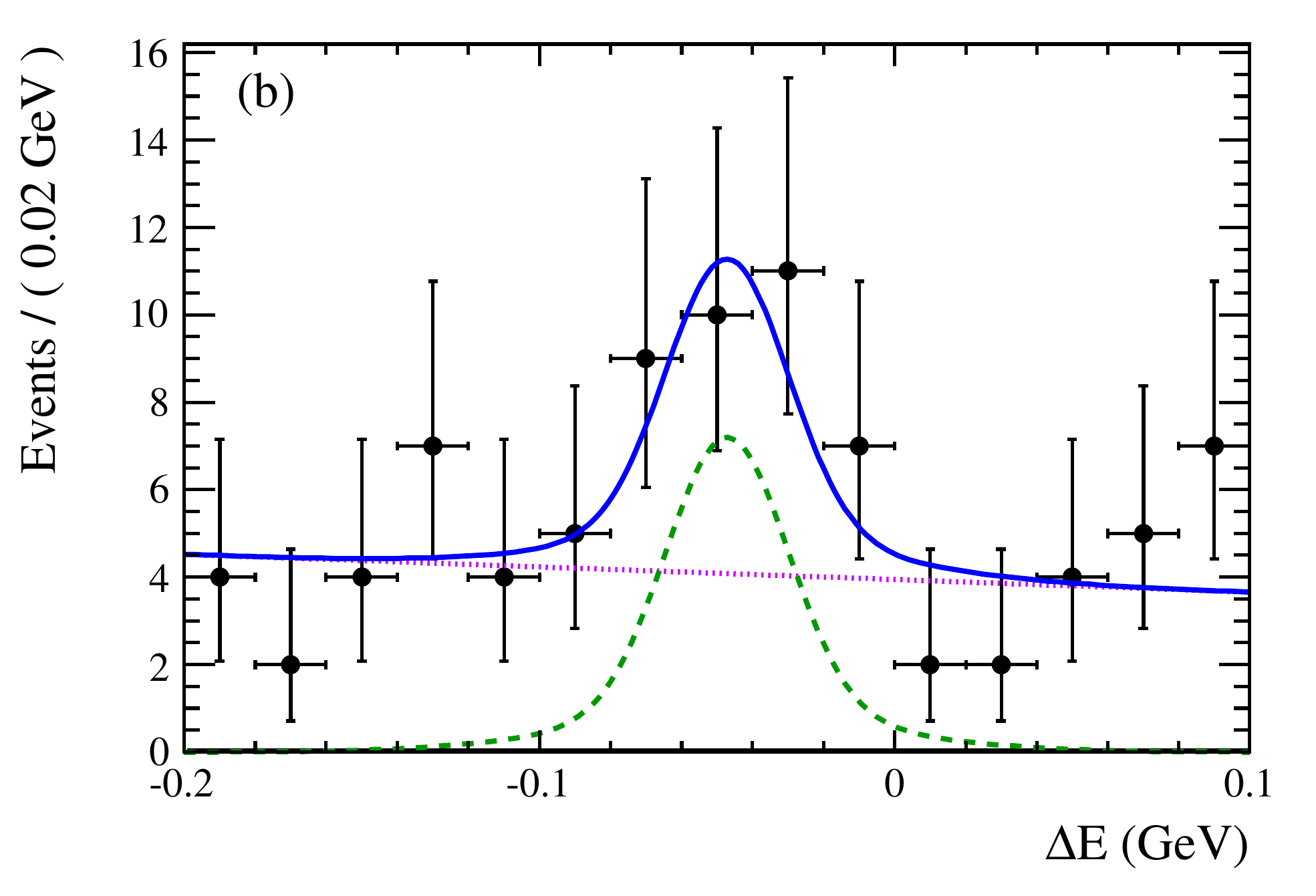}
  \includegraphics[width=0.32\textwidth]{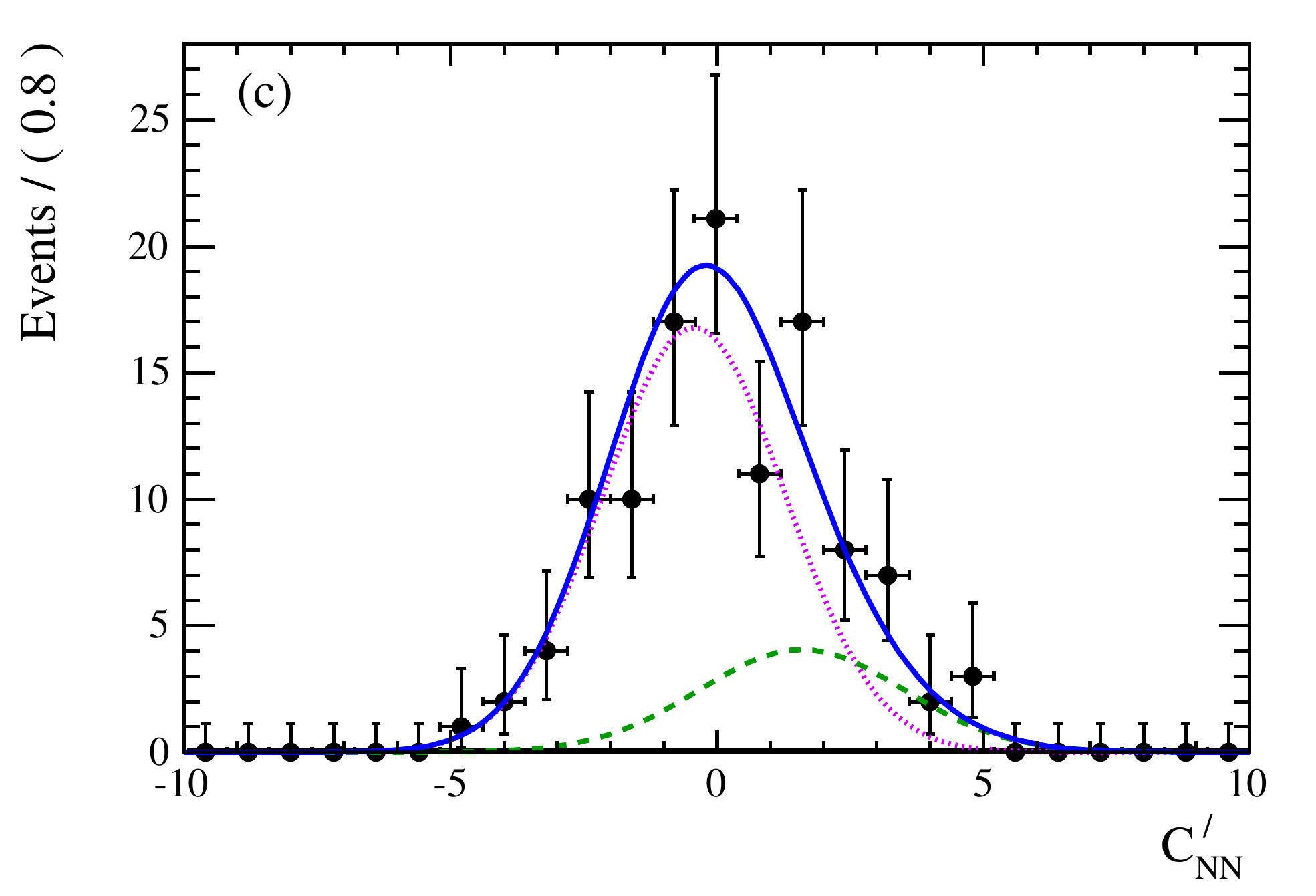}
\caption{\small  Projections of the 3D fit to the real data: 
(a) $M_{\rm bc}$ in $-0.11~{\rm GeV} <\Delta E < 0.02~{\rm GeV}$
and $C^{\prime}_{\rm NN}>0.5$;
(b) $\Delta E$ in $5.405~{\rm GeV}/c^{2} <M_{\rm bc}< 5.427~{\rm GeV}/c^{2}$
and $C^{\prime}_{\rm NN}>0.5$; and 
(c) $C^{\prime}_{\rm NN}$ in $5.405~{\rm GeV}/c^{2} <M_{\rm bc}< 5.427~{\rm GeV}/c^{2}$
and $-0.11~{\rm GeV} <\Delta E < 0.02~{\rm GeV}$. 
The points with  error bars are data, the (green) dashed curves
show the signal, (magenta) dotted curves show the continuum
background, and (blue) solid  curves show the total.  The three peaks in $M_{\rm bc}$ arise from 
$\Upsilon(5S)\to B_s^0\bar{B_s^0}, B_s^{*0}\bar{B_s^0}+B_s^0\bar{B}_s^{*0}$, and $B_s^{*0}\bar{B}_s^{*0}$ decays.
}
\label{fig:fig2}
\end{figure*}

Candidate $K^0$ mesons are reconstructed via the decay $K_S^0\to\pi^+\pi^-$ and require that the $\pi^+\pi^-$ invariant mass be within 12 MeV/$c^2$ of the nominal $K_S^0$ mass~\cite{PDG}. In order to extract the signal yield, we perform a three-dimensional (3D) unbinned maximum likelihood fit to the variables,  $M_{\rm bc}$,
$\Delta E$, and continuum suppression variable $C'_{\rm NN} = \ln\left(\frac{C_{\rm NN}-C^{\rm min}_{\rm NN}}
{C^{\rm max}_{\rm NN}-C_{\rm NN}}\right)$. We extract $29.0\,^{+8.5}_{-7.6}$ signal events
and $1095.0\,^{+33.9}_{-33.4}$ continuum background events.
Projections of the fit are shown in Fig.~\ref{fig:fig2}.
The branching fraction of the decay $B_s^0\rightarrow K^0\bar{K^0}$ is measured to be~\cite{Pal:2015ghq}
\begin{equation}
\mathcal{B}(B_s^0\rightarrow K^0\bar{K^0})=(19.6\,^{+5.8}_{-5.1}\,\pm1.0\,\pm2.0)\times10^{-6},
\end{equation}
where the first uncertainty is statistical, the second
is systematic, and the third reflects the uncertainty due to the 
total number of $B_s^0\bar{B_s^0}$ pairs. The significance of this result is 5.1 standard deviations, thus, our measurement constitutes the first observation of this decay.
This measured branching fraction  is in good agreement with the
SM predictions~\cite{SM-branching}, and it implies that the Belle II experiment~\cite{Abe:2010gxa} will
reconstruct over 1000 of these decays. Such a sample would allow for a much higher sensitivity search for new physics in this $b\to s$ penguin-dominated decay.

\section{Radiative $B_s^0$ decays}
\begin{figure}[htb]
\centering
\includegraphics[width=0.4\textwidth]{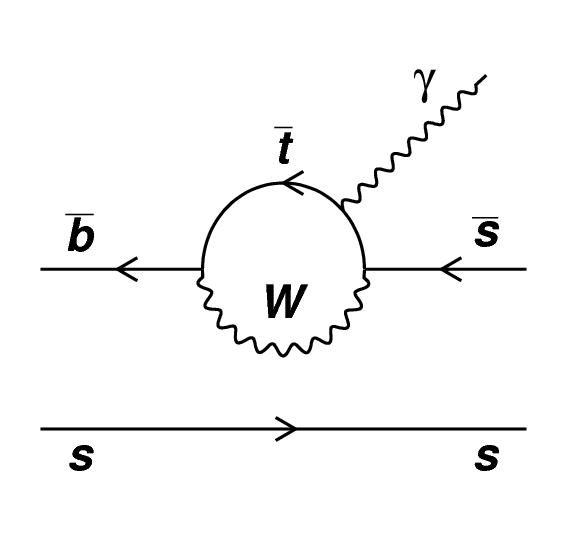}%
\includegraphics[width=0.4\textwidth]{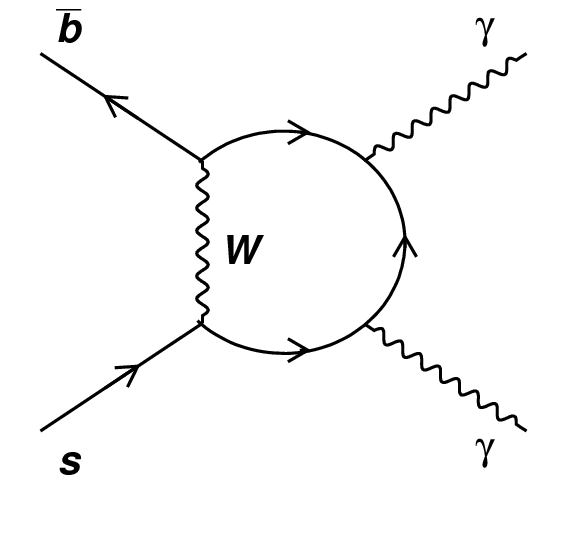}
\caption{\small Leading-order Feynman diagrams for  decays (left) $B_s^0\to\phi\gamma$ and (right) $B_s^0\to\gamma\gamma$.  }
\label{fig:feynman1}
\end{figure}

In the SM, the  decays $B_s^0\to\gamma\gamma$ and $B_s^0\to\phi\gamma$ are explained by the radiative transitions $b\to s\gamma\gamma$ and $b\to s\gamma$, respectively. 
The leading-order Feynman diagrams for these processes are shown in Fig.~\ref{fig:feynman1}.  First observation of the decay $B_s^0\to\phi\gamma$ was made by the Belle Collaboration using $23.6~{\rm fb^{-1}}$ 
of data collected at the $\Upsilon(5S)$ resonance and its branching fraction was measured to be $(5.7\,^{+2.2}_{-1.9})\times10^{-5}$~\cite{Wicht:2007ni}. The decay $B_s^0\to\gamma\gamma$ , on the other hand, has not been 
observed yet and  the  current upper limit on the branching fraction is $8.7\times10^{-6}$ at 90\% C.L.~\cite{Wicht:2007ni}. This is almost an order of magnitude larger than the range covered by the published theoretical calculations~\cite{ggth}.
New physics could enhance its branching fraction by more than an order of magnitude~\cite{Gemintern:2004bw}.

The results presented here are based on $121.4~{\rm fb^{-1}}$ recorded at the~$\Upsilon(5S)$. 
Candidate $\phi$ mesons are reconstructed via the decay $\phi\to K^+K^-$ and require that the $K^+K^-$ invariant mass be within 12 MeV/$c^2$ of the nominal $\phi$ mass~\cite{PDG}.
For $B_s^0\to\phi\gamma$  ($B_s^0\to\gamma\gamma$) decay, we perform a four-dimensional  (two-dimensional) unbinned maximum likelihood fit to the variables $M_{\rm bc}$, $\Delta E$, $C'_{\rm NN}$ and 
$\cos\theta_{\rm hel}$ ($M_{\rm bc}$ and $\Delta E$). The helicity angle $\theta_{\rm hel}$ is the angle between the $B_s^0$ and the $K^+$ evaluated in the $\phi$ rest frame.

\begin{figure}[htb]
\centering
\includegraphics[width=0.4\textwidth]{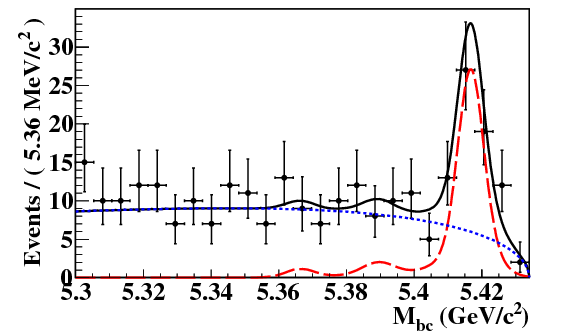}%
\includegraphics[width=0.4\textwidth]{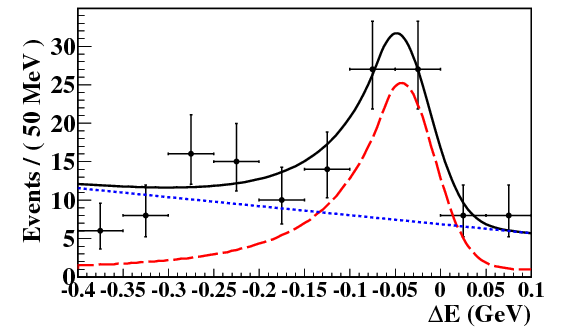}
\includegraphics[width=0.4\textwidth]{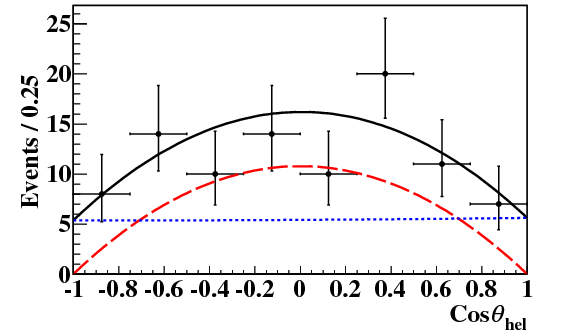}%
\includegraphics[width=0.4\textwidth]{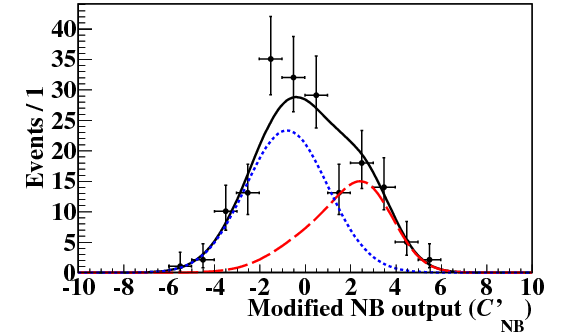}
\caption{\small Data fits for the $B_s^0\to\phi\gamma$ analysis. The projections are shown only for events inside the $B_s^*\bar{B_s^*}$ signal region except for the plotted variable. The $B_s^*\bar{B_s^*}$ signal region is defined as $M_{\rm bc}>5.4 ~{\rm GeV}/c^2, ~−0.2~{\rm GeV} <\Delta E< 0.02~{\rm GeV},~ |\cos\theta_{\rm hel}|< 0.8 ~\textrm{and}~ 0.0 <C'_{\rm NN}< 10.0$. The points with error bars represent the data, the solid black curve represents the total fit function, the red dashed (blue dotted) curve represents the signal (continuum background) contribution.  }
\label{fig:phigamma}
\end{figure}

We observe $91\,^{+14}_{-13}$ signal events in the $B_s^0\to\phi\gamma$ mode and the corresponding branching fraction is measured to be~\cite{Dutta:2014sxo}
\begin{equation}
\mathcal{B}(B_s^0\to\phi\gamma)=(36\,\pm5\,\pm3\,\pm6)\times10^{-6},
\end{equation}
where the first uncertainty is statistical, the second
is systematic, and the third reflects the uncertainty due to the fraction of $B_s^{(*)}\bar{B}_s^{(*)}$ in $b\bar{b}$ events. Fit projections are shown in Fig.~\ref{fig:phigamma}. This improved result
supersedes our earlier measurement~\cite{Wicht:2007ni} and is consistent with the recent LHCb's measurement~\cite{Aaij:2012ita} .

\begin{figure}[htb]
\centering
\includegraphics[width=0.4\textwidth]{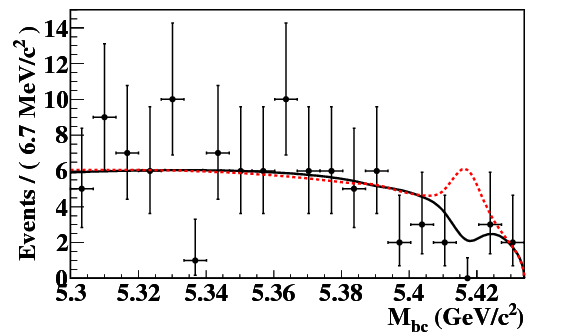}%
\includegraphics[width=0.4\textwidth]{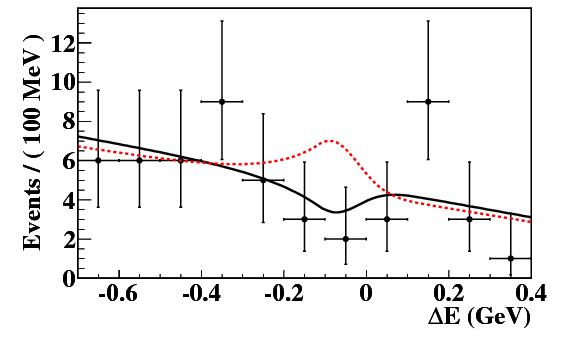}
\caption{\small Data fits for the $B_s^0\to\gamma\gamma$ analysis. The projections are shown only for events inside the $B_s^*\bar{B_s^*}$ signal region except for the plotted variable. The $B_s^*\bar{B_s^*}$ signal region is defined as $M_{\rm bc}>5.4 ~{\rm GeV}/c^2 ~\textrm{and}~ ~−0.3~{\rm GeV} <\Delta E< 0.05~{\rm GeV}$. The points with error bars represent the data, the solid black curve represents the total fit function, the red dashed (blue dotted) curve represents the signal (continuum background) contribution.  }
\label{fig:gammagamma}
\end{figure}

We see no significant signal in the $B_s^0\to\gamma\gamma$ mode and we extract an upper limit at 90\% C.L. of ~\cite{Dutta:2014sxo}
\begin{equation}
\mathcal{B}(B_s^0\to\gamma\gamma)<3.1\times10^{-6}.
\end{equation}
This result represent an improvement by a factor of about 3 over the previous best measurement~\cite{Wicht:2007ni}. Fit projections are shown in Fig.~\ref{fig:gammagamma}.

\section{Conclusions}
Using the full set of Belle data collected at $\Upsilon(5S)$ resonance, recent measurements of charmless
hadronic and radiative $B_s^0$ decays are presented.  Our measurement of
$B_s^0\to K^0\bar{K^0}$
branching fraction
constitutes the  first observation of the decay. This is the first observation of a charmless $B_s^0$ decay involving only neutral hadrons. 

\section*{Acknowledgements}
The author thanks the organizers of XIII International Conference on Heavy Quarks and Leptons for excellent hospitality and for assembling a nice scientific program.  
This work is
supported by the U.S. Department of Energy.



\begin{thebibliography}{99}
\bibitem{charge-conjugate}
Unless stated otherwise, charge-conjugate modes are implicitly included.
\end{thebibliography}
\end{document}